\documentclass[aps,prl,twocolumn,superscriptaddress,10pt]{revtex4-1}
\usepackage{graphicx}
\usepackage[normalem]{ulem}
\usepackage{amsmath}
\usepackage{hyperref}
\hypersetup{urlcolor=blue, colorlinks=false} 
\usepackage{color}
\usepackage{bm}
\usepackage{braket}

\definecolor{orange}{rgb}{1,0.5,0}

\begin{document}

\title{Relativistic response and novel spin-charge plasmon at the Tl/Si(111) surface}

\author{Jon Lafuente-Bartolome}
\affiliation{Materia Kondentsatuaren Fisika Saila, Euskal Herriko Unibertsitatea UPV/EHU, 644 Postakutxatila, 48080 Bilbao, Basque Country, Spain}
\affiliation{Donostia International Physics Center (DIPC), Paseo Manuel de Lardizabal 4, 20018 Donostia-San 
Sebastian, Basque Country, Spain}
\author{Idoia G. Gurtubay}
\affiliation{Materia Kondentsatuaren Fisika Saila, Euskal Herriko Unibertsitatea UPV/EHU, 644 Postakutxatila, 48080 Bilbao, Basque Country, Spain}
\affiliation{Donostia International Physics Center (DIPC), Paseo Manuel de Lardizabal 4, 20018 Donostia-San 
Sebastian, Basque Country, Spain}
\author{Asier Eiguren}
\affiliation{Materia Kondentsatuaren Fisika Saila, Euskal Herriko Unibertsitatea UPV/EHU, 644 Postakutxatila, 48080 Bilbao, Basque Country, Spain}
\affiliation{Donostia International Physics Center (DIPC), Paseo Manuel de Lardizabal 4, 20018 Donostia-San 
Sebastian, Basque Country, Spain}

\date{\today}

\begin{abstract}
We present a comprehensive \textit{ab initio} analysis of the spin-charge correlations at the Tl/Si(111) surface, 
where the spin-orbit interaction is so strong that a detailed treatment of the non-collinear electron spin appears decisive for 
the correct description of the response properties. The relativistic limit enforces a unified treatment of the spin and charge densities as a four-vector, and the response function acquires then a 4$\times$4 tensor structure. Our all-electron implementation allows to
resolve the real space structure of the possible collective modes, and demonstrates the emergence of a novel collective excitation combining transverse-spin and ordinary charge oscillations of a similar order of magnitude, whose spin character is strongly enhanced as we approach the $q$$\rightarrow$0 momentum limit.

\end{abstract}

\maketitle

Understanding the role of the spin on the dynamics of many-electron systems is of paramount importance. Contrary to conventional electronics, primarily based on the charge property of the electron states, controlling quasi-particles or emergent many-body collective modes focusing on the spin state has become a very challenging but realistic possibility \cite{RevModPhys.76.323}. 
A key simplifying idea or approximation of contemporary many-body physics has been the concept of the collective mode \cite{PhysRev.85.338,Pines}. Among others, the so-called plasmon and magnon states have been historically considered as the collective modes associated with the real space oscillations of charge and spin densities, respectively. Even without considering the spin-orbit interaction, three decades of systematic quantum-mechanical studies of response functions in real surfaces have led to the prediction and discovery of unexpected collective phenomena, such as the acoustic surface plasmon \cite{0295-5075-66-2-260,Diaconescu2007}. However, relativity is known to introduce a peculiar interplay between the electron charge and spin, and it has been proven to produce well defined spin textures even in nominally non-magnetic materials \cite{1367-2630-17-5-050202}. The last decade has been characterized by the emergence of several entirely new research fields focusing on the study of the role of the electron spin in presence of relativistic corrections and its associated topological properties \cite{soumyanarayanan2016emergent,RevModPhys.82.3045,RevModPhys.83.1057}. 
This is so because of the wide variety of fascinating electromagnetic and transport properties shown by these materials, such as the Edelstein \cite{Edelstein,PhysRevLett.93.176601} or the spin Hall \cite{1971JETPL..13..467D,Kato1910} effects, which have triggered a justified expectation about possible applications.

Recently, several model theoretical studies considering the Rashba or Dresselhaus spin-orbit interaction 
in ideal homogeneous two-dimensional systems have predicted the presence of novel types of collective excitations induced by the spin-orbit coupling, such as the so-called chiral spin waves \cite{PhysRevLett.109.227201,PhysRevB.91.035106,PhysRevB.79.205305,PhysRevLett.104.116401}. 
A particular mention to Ref.\cite{PhysRevLett.104.116401} is in order, as in this work the authors 
predicted the presence of a $\omega_{q}\sim\sqrt{q}$ dispersing mode in a helically polarized ideal Dirac system, 
showing that this mode is constituted by both spin and ordinary charge oscillations.
Along this line, metal surfaces and semiconductor/heavy-element overlayers holding spin-polarized quasi-2D surface states may be potential systems for showing mixed spin and charge collective excitations.  
However, a thorough \textit{ab initio} analysis of such excitations in real materials incorporating the details of complex band structures and strong spin-orbit coupling is still lacking. In this respect, the Tl/Si(111) surface appears as an excellent illustrative example due to the exceptionally strong spin-orbit induced spin-splitting of its surface states ($\sim$0.5~eV). Moreover, these states show a characteristic spin texture composed by a Rashba-like chiral pattern close to the $\bar{\Gamma}$ point and  a complete surface-perpendicular spin polarization at the high symmetry point(s) $\bar{K}$ ($\bar{K}'$) \cite{PhysRevB.84.125435,PhysRevLett.102.096805,PhysRevLett.111.176402}.

In this article, we present a first-principles analysis of the 
relativistic response properties of the Tl/Si(111) surface system, 
incorporating the spinor structure of the electron wave function as well as the fine structure details of the electron bands and local field effects. 
The relativistic limit induces an intimate
coupling between electron spin and charge densities, and it seems natural to introduce a four-density vector, ${\bf n}^{\mu}=(\rho,m_{x},m_{y},m_{z})$, describing the ordinary charge density together with the
three possible Cartesian components of the spin density.
Thus, our response functions are represented by a $4\times4$ tensor and the 16 components encode all the possible charge/spin density correlations in the presence of an external electromagnetic field.
We will show that owing to the particular band structure and spin texture of the surface states present in this system, a novel low energy collective mode emerges, which is composed by coexisting charge and spin density oscillations. Our first-principles scheme allows to resolve the real space configuration of the spin/charge character of such oscillations.

Within the framework of Spin Density Functional Theory (SDFT), Kohn-Sham (KS) wave functions are generalized by a two component spinor at each $\textbf{k}$ point in the first Brillouin zone,
\begin{equation}\label{eq:spinor}
\Psi_{n,\textbf{k}}({\bf r})= 
\begin{pmatrix}
\varphi_{n,\textbf{k}}^{(\uparrow)}({\bf r}) \\
\varphi_{n,\textbf{k}}^{(\downarrow)}({\bf r})
\end{pmatrix}~,
\end{equation}
where $\varphi_{n,\textbf{k}}^{(\uparrow)}({\bf r})$ and $\varphi_{n,\textbf{k}}^{(\downarrow)}({\bf r})$ represent the up/down components for a given direction. The components of the spinor wave function satisfy a set of two coupled KS equations \cite{0022-3719-5-13-012} in which the effective scalar potential is replaced by a spin-dependent one and the ordinary electron density becomes a four component spin-density matrix, 
\begin{equation}\label{eq:densitymatrix}
 {\bf n}^{\alpha\beta}({\bf r})=\sum_{n,{\bf k}}^{\text{occ}} \varphi_{n,{\bf k}}^{\alpha}({\bf r})(\varphi^{\beta}_{n,{\bf k}}({\bf r}))^{*}~.
\end{equation}

As for linear response theory, the generalization of the non-interacting density-density response function leads, as mentioned above, to a 4$\times$4 response matrix and is directly accesible by considering the ground-state KS spinors of Eq.(\ref{eq:spinor}) by \cite{PhysRev.126.413,PhysRev.129.62},
\begin{equation}\label{eq:chi_0}
\begin{split}
\boldsymbol{\chi}&_{\text{KS}}^{\alpha\beta\alpha'\beta',{\bf GG'}}({\bf q},\omega)=\frac{1}{\Omega}\sum_{{\bf k}}^{1BZ}\sum_{n,m} ~ \frac{(f_{n{\bf k}}-f_{m{\bf k+q}})}{\omega + (\epsilon_{n{\bf k}}-\epsilon_{m{\bf k+q}})+i\eta} \\
&\langle \varphi^{\beta}_{n,{\bf k}} | e^{-i({\bf q+G) \cdot r}} | \varphi^{\alpha}_{m,{\bf k+q}}\rangle \langle \varphi^{\alpha'}_{m,{\bf k+q}} | e^{i({\bf q+G') \cdot r}} | \varphi^{\beta'}_{n,{\bf k}}\rangle ~ .
\end{split}
\end{equation}
Following the TDDFT scheme \cite{PhysRevLett.52.997,PhysRevLett.76.1212}, the full interacting response tensor $\boldsymbol{\chi}$ can then be obtained by means of the Dyson equation, which formally reads,
\begin{equation}\label{eq:dyson}
\boldsymbol{\chi} = [{\bf1} - \boldsymbol{\chi}_{\text{KS}}~{\bf F}_\text{xc}]^{-1} ~ \boldsymbol{\chi}_{\text{KS}}~.
\end{equation}
In principle, ${\bf F}_\text{xc}$ includes both the Coulomb interaction and the exchange-correlation kernel, and the inhomogeneities in real space (so-called local field effects) are taken into account by the non-diagonal ${\bf G}\neq{\bf G'}$ elements in Eq.(\ref{eq:chi_0}).

In Eq.(\ref{eq:chi_0}) we express the generalized response matrix in the spinor basis, but a clearer physical interpretation is obtained by means of the more usual tensor representation considering the Pauli tetrad basis $\boldsymbol{\sigma}^{\mu} \equiv (\boldsymbol{\sigma}^{0},\boldsymbol{\sigma}^{x},\boldsymbol{\sigma}^{y},\boldsymbol{\sigma}^{z})$,
\begin{eqnarray}\label{eq:spin_tetrad_transf}
%
%
%
{\bf n}^{\mu} &=& \frac{1}{2}\sum_{\alpha\beta} \boldsymbol{\sigma}^{\mu}_{\alpha\beta} {\bf n}^{\beta\alpha}~, \\
\boldsymbol{\chi}^{\mu\nu} &=& \frac{1}{4}\sum_{\alpha\beta\alpha'\beta'} \boldsymbol{\sigma}^{\mu}_{\beta\alpha} ~ \boldsymbol{\chi}^{\alpha\beta\alpha'\beta'}   \boldsymbol{\sigma}^{\nu}_{\alpha'\beta'}~,
\end{eqnarray}
where ${\bf n}^{\mu}$ represents the four-component density vector ${\bf n}^{\mu}=(\rho,m_{x},m_{y},m_{z})$. In this way, we arrive at the generalized linear spin-charge density response equation,
\begin{equation}\label{eq:lresponse}
\delta {\bf n}^{\mu}(\textbf{r},\omega) = \sum_{\nu} ~ \int d^{3}r' \boldsymbol{\chi}^{\mu\nu}(\textbf{r},\textbf{r}',\omega) ~ \delta \boldsymbol{\phi}^{\nu(\text{ext})}(\textbf{r}',\omega) ~,
\end{equation}
which relates the induced charge and spin densities, and the external electromagnetic field, $\delta\boldsymbol{\phi}^{\nu(\text{ext})}=(\delta V_{0}^{(\text{ext})},\delta H_{x}^{(\text{ext})},\delta H_{y}^{(\text{ext})},\delta H_{z}^{(\text{ext})})$.


\begin{figure}[t]
\centering{
{\includegraphics[width=0.95\columnwidth,angle=0]{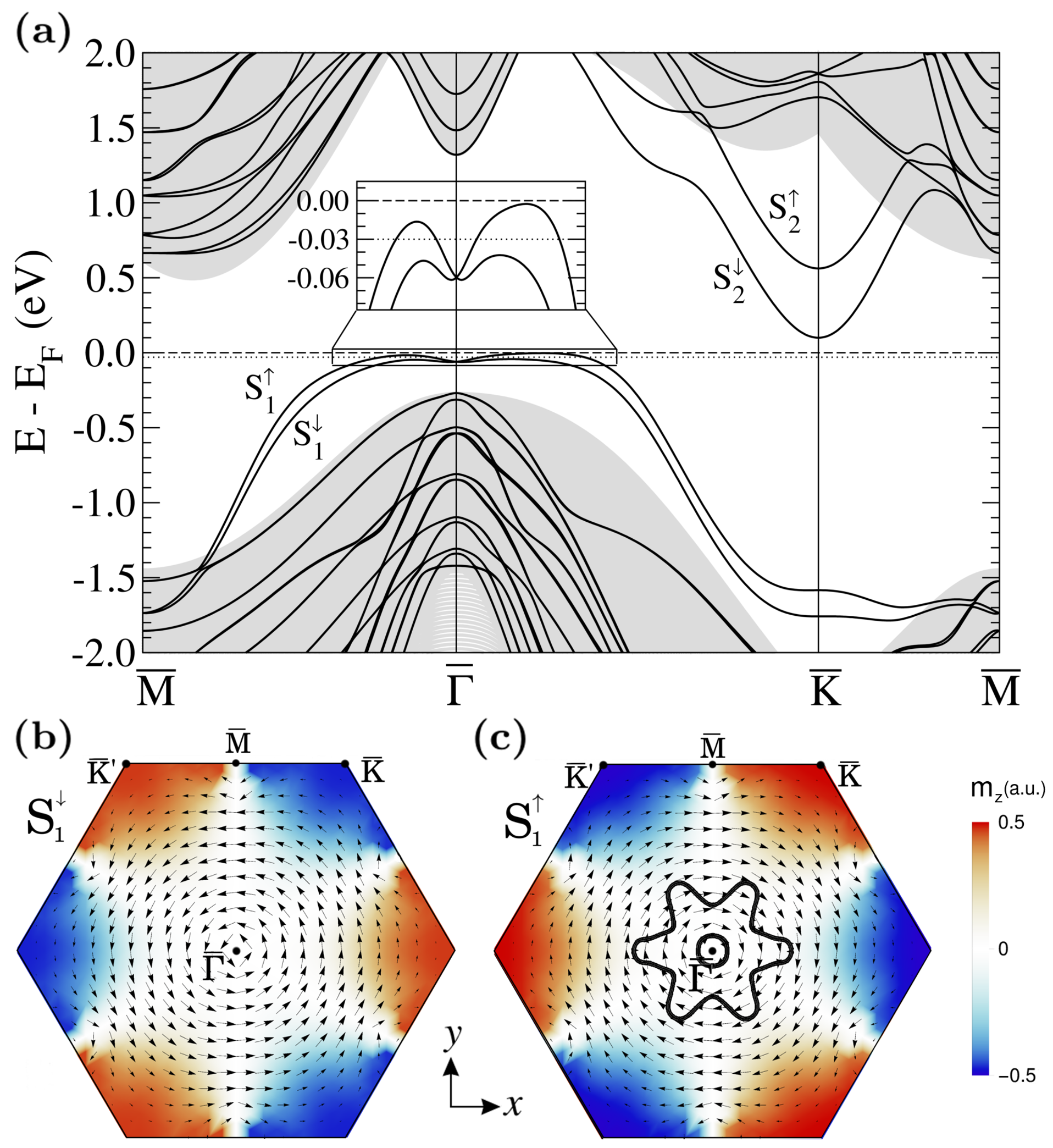}}
}
\begin{center}
\caption {Ground state electronic structure of the Tl/Si(111) surface. (a) Calculated band structure. Energies are given with respect to Fermi energy, which is represented by the dashed line. The dotted line shows the shifted Fermi level used in the response function calculations. The light gray background represents the bulk band projection. The inset shows a zoom of the $\text{S}_1^{\downarrow}$ and $\text{S}_1^{\uparrow}$ surface bands near the $\bar{\Gamma}$ point. (b) and (c) Momentum dependent spin polarization of the two occupied surface states $\text{S}_1^{\downarrow}$ and $\text{S}_1^{\uparrow}$, respectively, over the whole first surface Brillouin zone. Arrows represent the in-plane spin polarization components, whereas the color code represents the out-of-plane spin polarization component. The Fermi contour corresponding to the Fermi level shifted by $-0.03$~eV is represented by the black solid line in (c).}\label{fig:fig1}
\end{center}
\end{figure}



\begin{figure*}[t]
\centering{
\includegraphics[width=2.0\columnwidth,angle=0]{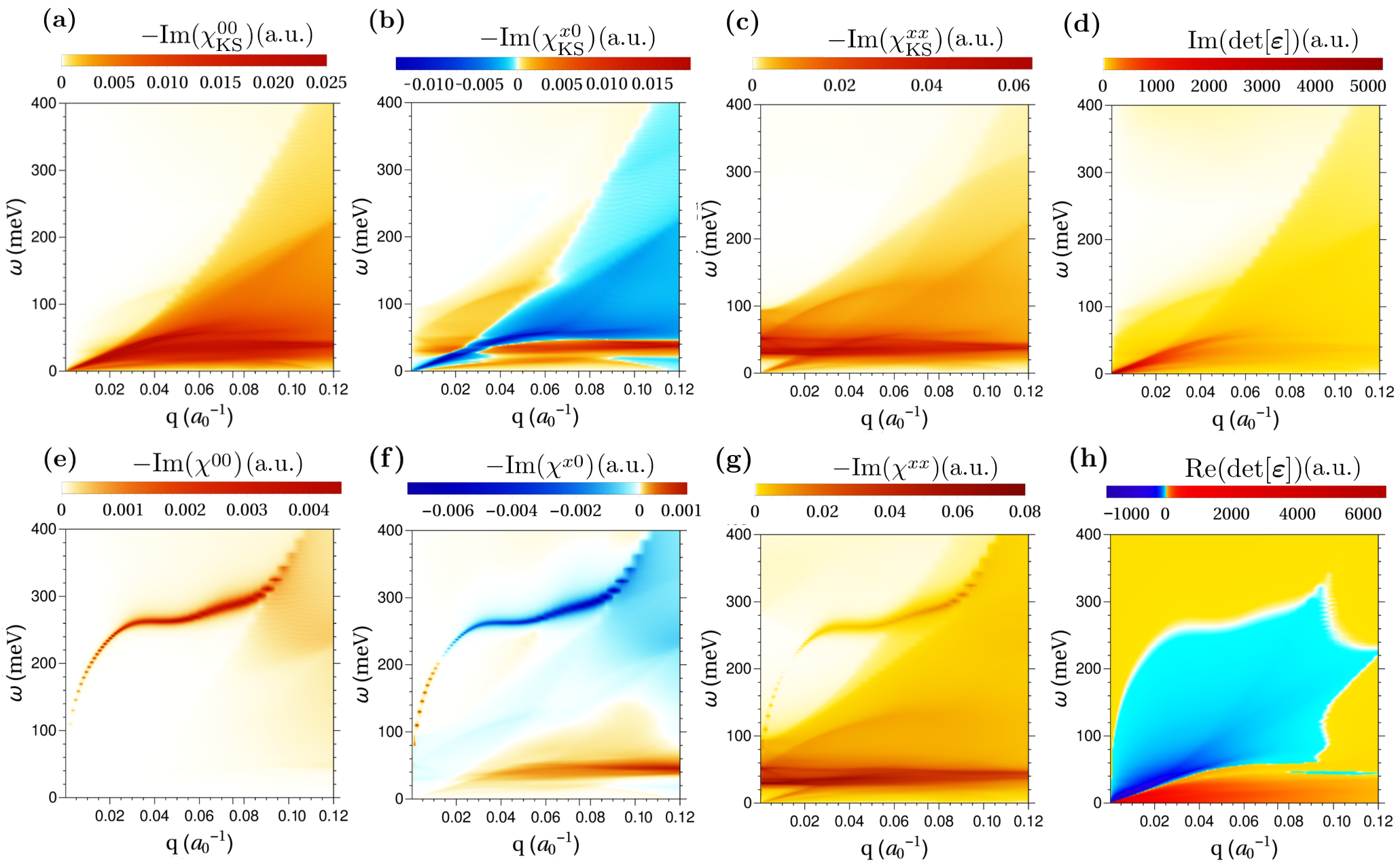}
}
\begin{center}
\caption {
  Selected components of the calculated spin-charge response tensor for ${\bf q}$ along the $\bar{\Gamma}-\bar{\text{M}}$ direction (corresponding to $y$ in our coordinate system). Panels (a)-(c) show the calculated charge/charge, transverse-spin/charge and transverse-spin/transverse-spin
  response functions, respectively, for the non-interacting case (as in Eq.(\ref{eq:chi_0})). Panels (e)-(g) show their interacting counterparts, calculated from Eq.(\ref{eq:dyson}). Panels (d) and (h) represent the imaginary and real parts of the determinant of the spin-charge dielectric response tensor (see Eq.(\ref{eq:epsilon})), respectively, which are relevant for determining the presence of possible collective modes and their real space details (see Eqs.(\ref{eq:extpot-scpot})-(\ref{eq:eigenproblem}) and discussion therein).
}
\label{fig:fig2}
\end{center}
\end{figure*}


The Tl/Si(111) surface was simulated considering a slab system consisting of 10 silicon layers with a thallium adlayer and a vacuum space of $51a_{0}$ between the repeated slabs. On the other side of the slab a hydrogen adlayer was introduced in order to saturate the dangling bonds. Ground state calculations were performed using the all-electron LAPW method \cite{elk}, considering a 24$\times$24$\times$1 Monckhorst-Pack grid \cite{PhysRevB.13.5188} and the non-collinear LSDA approximation for the exchange-correlation energy \cite{PhysRevB.45.13244}. Spin-orbit interaction has been included self-consistently in all the ground state calculations.

Fig.\ref{fig:fig1} shows the essential information about the electronic band structure and spin polarization of the Tl/Si(111) surface, as obtained by means of our relativistic ground state calculations. In panel (a), the solid black lines correspond to the slab bands, while the continuous grey background represents the projected band structure of bulk silicon. Panels (b) and (c) display the calculated momentum-dependent spin textures of the occupied surface bands $\text{S}_1^{\downarrow}$ and $\text{S}_1^{\uparrow}$, defined as the expectation value of the Pauli matrices, 
\begin{equation}\label{eq:spinpol}
{\bf m}_n({\bf k})= \frac{1}{\Omega} \int d^{3}r \Psi_{n{\bf k}}^{\dagger}({\bf r}) \boldsymbol{\sigma} \Psi_{n{\bf k}}({\bf r})~,
\end{equation}
where $\Omega$ represents the volume of the unit-cell. Our results compare well with previous SDFT calculations based on the pseudopotential method \cite{PhysRevB.84.125435}, as well as with angle and spin resolved photoemission experiments \cite{PhysRevLett.102.096805,PhysRevLett.111.176402}. 
The Tl/Si(111) surface states show a rich non-collinear spinor structure, and a  moderate hole-doping of 30~meV (see dotted-line in the inset of Fig.\ref{fig:fig1}(a)) results in a fully spin-polarized Fermi surface with chiral spin texture, 
with only  the upper spin-split subband $\text{S}^{\uparrow}_{1}$ crossing the Fermi level.
Furthermore, since the splitting of bands $\text{S}^{\uparrow}_{1}$ and $\text{S}^{\downarrow}_{1}$ remains almost constant in the vicinity of the $\bar{\Gamma}$ point,  the band structure of this surface system near this point deviates substantially from that of the pure Rashba-like systems.

Fig.\ref{fig:fig2} presents our results for the generalized spin-charge density response tensor of the hole-doped Tl/Si(111) surface. For the sake of simplicity, we have focused on a momentum ${\bf q}$ along the $\bar{\Gamma}-\bar{\text{M}}$ direction, which corresponds to the $y$ axis in our coordinate system. Therefore, from now on we refer to the coordinate $x$ as the transverse direction. Panels (a)-(c) of Fig.\ref{fig:fig2} show the calculated macroscopic contributions ($\boldsymbol{\chi}_{\text{KS}}^{\mu\nu,{\bf G}=0{\bf G'}=0}({\bf q},\omega)$) of the non-interacting charge/charge, transverse-spin/charge, and transverse-spin/transverse-spin responses, respectively, while panels (e)-(g) of Fig.\ref{fig:fig2} show their full interacting counterparts (see Eq.(\ref{eq:lresponse})) \footnote{We only show these three components of the response tensor for being the most meaningful ones for this particular direction of the momentum transfer vector (all the sixteen components of the response tensor are displayed in Ref.\cite{Suppl.Mat.})}. The non-interacting 4$\times$4 component response function has been obtained evaluating the summation of Eq.(\ref{eq:chi_0}) over a dense 840$\times$840 \textbf{k}-point grid, in which all the \textbf{k}-dependent elements have been interpolated using the Wannier-interpolation technique \cite{PhysRevB.56.12847,PhysRevB.65.035109}. This procedure allows to achieve converged results considering a damping parameter as fine as $\eta=1$~meV, which permits to obtain a smooth \textbf{q}-dependent map of the response functions \cite{Suppl.Mat.}. As a next step, the interacting response has been obtained by direct inversion of Eq.(\ref{eq:dyson}) where we keep the local field effects. We use the LSDA approximation of the exchange-correlation kernel \cite{PhysRevB.45.13244}, and we consider a truncation of the Coulomb potential in the direction perpendicular to the surface in order to avoid artificial interaction between the slabs \cite{PhysRevB.73.205119}.

The intraband single-particle excitation continuum can be noticed in the three components of the non-interacting response tensor (see Fig.\ref{fig:fig2}(a-c)) \cite{Pines}. Moreover, the interband excitation continuum (also called ``Rashba'' continuum \cite{PhysRevB.91.035106}) is also visible in the Im($\chi_{\text{KS}}^{x0}$) and Im($\chi_{\text{KS}}^{xx}$) components. Interestingly, we find that interband transitions carry a change of sign in Im($\chi_{\text{KS}}^{x0}$) with respect to the intraband transitions, an effect originating from the opposite spin orientation of the two spin-split subbands. These three panels define the regions of the ({\bf q},$\omega$) space where the possible collective modes of the system may suffer from damping due to single-particle excitations.

Turning back to the full interacting response, a prominent peak on the charge/charge response component (Im($\chi^{00}$)) is observable in Fig.\ref{fig:fig2}(e), which lies well above the single-particle excitation continuum up to $|{\bf q}|\sim0.08~\text{a}_{0}^{-1}$, clearly resembling the $\omega_{q}\sim\sqrt{q}$ dispersion of a quasi-2D charge plasmon \cite{PhysRevLett.86.5747,PhysRevB.78.155402}. Noteworthy, the spin-charge interplay in the response introduced by the spin-texture of the surface states becomes manifest when we evaluate Im($\chi^{x0}$) and Im($\chi^{xx}$), as in both of these functions a similar peak is observed with exactly the same dispersion (see Fig.\ref{fig:fig2}(f-g)), together with the single-particle excitation background in the case of Im($\chi^{xx}$). 
As for Im($\chi^{x0}$), we observe a sign change and an intensity enhancement for low values of {\bf q}. This effect comes from the aforementioned sign change of the Im($\chi_{\text{KS}}^{x0}$) function, and is in principle also present in conventional pure Rashba-like systems. However, when considering the \textit{ab initio} response of a real surface such as Tl/Si(111) new features arise. The almost constant splitting between the subbands near $\Gamma$ (see inset of Fig.1(a)), makes the interband continuum to remain at low energies over the considered momentum range, allowing for a well-defined collective excitation -free from decaying into single-particle excitations- in the region of interest. In addition, for larger momentum transfers, another almost constantly dispersive peak starting at $|{\bf q}|\sim0.03~\text{a}_{0}^{-1}$ appears in Im($\chi^{x0}$) (see Fig.2(f)), coming from the low-energy interband transitions as well. For the sake of conciseness, from now on we will focus on the $\omega_{q}\sim\sqrt{q}$ dispersing mode.


\begin{figure}[t]
\centering{
\includegraphics[width=0.8\columnwidth,angle=0]{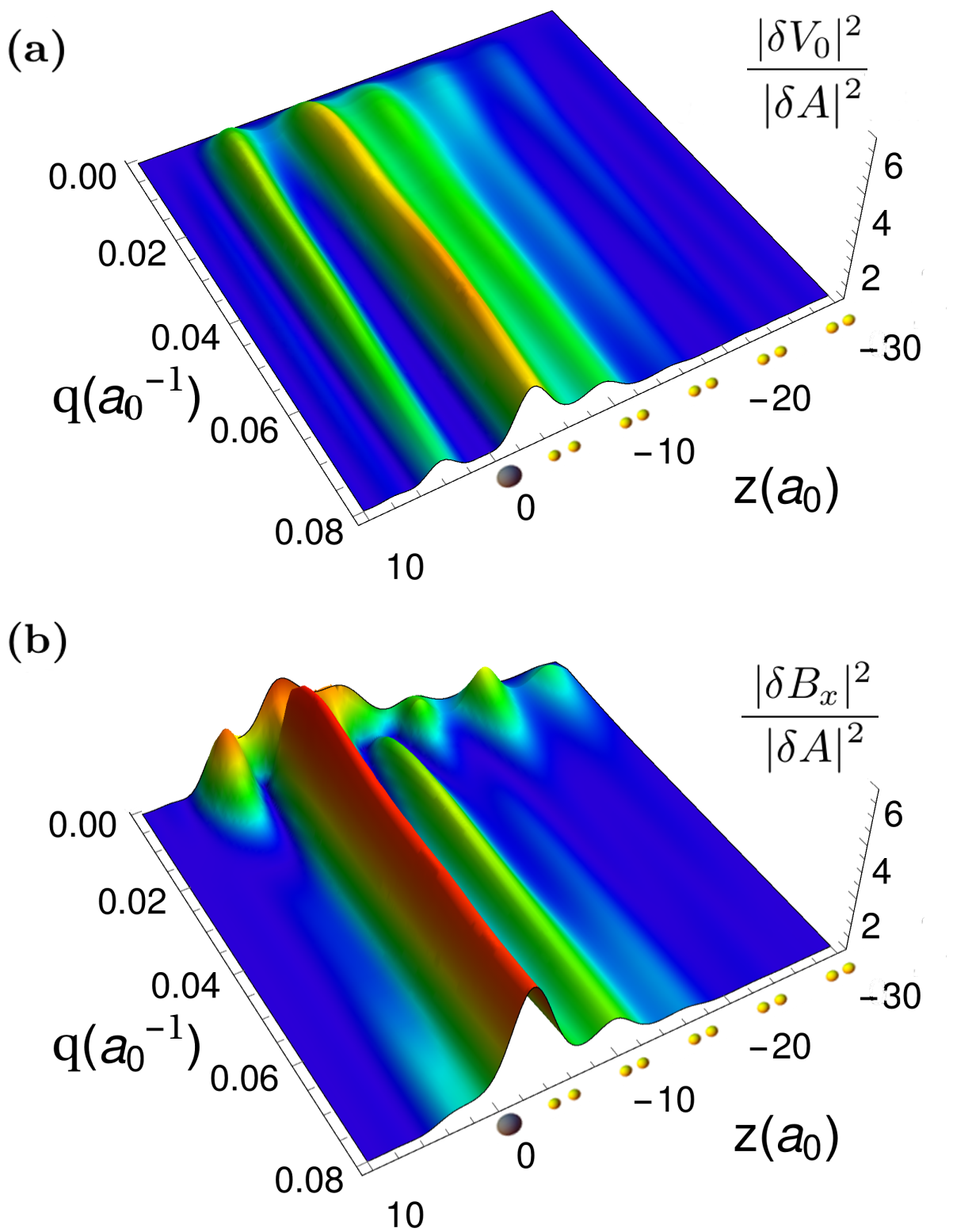}
}
\begin{center}
\caption {Real space configuration and ${\bf q}$ dependece of the coupled spin-charge collective oscillation at the Tl/Si(111) surface. (a) Magnitude of the induced charge potential oscillation and (b) induced transverse magnetic field oscillation, where $|\delta A|^{2} = \int_{cell} d^{3}r \sum_{\mu} (\delta\boldsymbol{\phi}({\bf r})^{\mu})^{*}\delta\boldsymbol{\phi}({\bf r})^{\mu}$. The $z$ coordinate corresponds to the direction perpendicular to the surface, with negative values indicating penetration into the bulk. Orientative positions of the first thallium and silicon atomic layers are represented by big gray and small yellow spheres, respectively.}\label{fig:fig3}
\end{center}
\end{figure}


In order to gain further insight in the real space details of this excitation, we compute the spin-charge dielectric response tensor $\boldsymbol{\varepsilon}$,
\begin{equation}\label{eq:epsilon}
 \boldsymbol{\varepsilon} = [{\bf 1} + \boldsymbol{\chi}~{\bf F_{\text{xc}}}]^{-1} = {\bf 1} - \boldsymbol{\chi}_{\text{KS}} ~ {\bf F_{\text{xc}}}~.
\end{equation}
This tensor relates the -four component- self-consistent potential ($\delta\boldsymbol{\phi}^{\text{sc}}=\delta\boldsymbol{\phi}^{\text{ext}} + \delta\boldsymbol{\phi}^{\text{ind}}$) and the external potential ($\delta\boldsymbol{\phi}^{\text{ext}}$),
\begin{equation}\label{eq:extpot-scpot}
 \delta\boldsymbol{\phi}^{\mu,{\bf G}(\text{ext})} = \sum_{\nu}\sum_{{\bf G'}} \boldsymbol{\varepsilon}^{\mu\nu,{\bf G}{\bf G'}} \delta\boldsymbol{\phi}^{\nu,{\bf G'}(\text{sc})}~.
\end{equation}
We ask for the self-sustained ($\delta\boldsymbol{\phi}^{\text{ext}}=0$) oscillations to fulfill the following condition,
\begin{equation}\label{eq:plasmoncond}
 \text{det}[\boldsymbol{\varepsilon}^{\mu\nu,{\bf G}{\bf G'}}({\bf q},\omega)]=0~.
\end{equation}
Above, the determinant has to be evaluated accounting for both the space and the spin degrees of freedom, and therefore, the dimension of the problem becomes 16 ($4$$\times$$4$) times larger than the scalar case. We can express Eq.(\ref{eq:extpot-scpot}) as an eigenvalue equation (in a similar way as in Ref.\cite{PhysRevB.86.245129} but including spin),
%
\begin{equation}\label{eq:eigenproblem}
 \sum_{\nu}\sum_{{\bf G'}} \boldsymbol{\varepsilon}^{\mu\nu,{\bf G}{\bf G'}}(\textbf{q},\omega) ~ \delta\boldsymbol{\phi}_{i}^{\nu,{\bf G'}}(\textbf{q},\omega) = \epsilon_{i}(\textbf{q},\omega) ~ \delta\boldsymbol{\phi}_{i}^{\mu,{\bf G}}(\textbf{q},\omega)~,
\end{equation}
so that the condition of Eq.(\ref{eq:plasmoncond}) is satisfied for the solution of Eq.(\ref{eq:eigenproblem}) with a vanishing eigenvalue ($\epsilon_{i}(\textbf{q},\omega)=0$). This procedure allows to resolve the spatial dependence and the mixed spin-charge character of the excitation  \cite{Suppl.Mat.}. We show in panels (d) and (h) of Fig.\ref{fig:fig2} the calculated imaginary and real parts of $\text{det}[\boldsymbol{\varepsilon}^{\mu\nu,{\bf G}{\bf G'}}]$, respectively. We can recognize the peaks in Im($\chi^{00}$), Im($\chi^{x0}$) and Im($\chi^{xx}$) as zeros of the function Re(det[$\boldsymbol{\varepsilon}^{\mu\nu,{\bf G}{\bf G'}}$]) which lie in regions with vanishingly small Im(det[$\boldsymbol{\varepsilon}^{\mu\nu,{\bf G}{\bf G'}}$]), thus identifying the excitation as a well definend self-sustained collective oscillation.

Fig. (\ref{fig:fig3}) shows the real space structure of the self-sustained oscillation as a function of ${\bf q}$ and $z$, being the latter the real space coordinate perpendicular to the surface.
The ordinary charge part ($\delta V_{0}$) and transverse-magnetic component ($\delta B_{x}$) of the oscillation 
are represented in panels (a) and (b), respectively. The longitudinal ($\delta B_{y}$) and surface perpendicular ($\delta B_{z}$) magnetic components are negligible in comparison, and are shown in Ref.\cite{Suppl.Mat.}.
For the sake of simplicity, we keep only the $z$ dependence by averaging the amplitudes in the surface-plane directions. 
The quasi-2D character of the mode is confirmed as both components remain localized within the first five atomic layers ($\sim$12~a$_0$) 
close to the surface area.
Most importantly, this figure reveals that the amplitude of the transverse-magnetic component is of a similar order of magnitude and even larger than the amplitude of the charge part over the considered momentum range. 
We also observe that the real space configuration of this mode 
is almost independent of the momentum except for the ${\bf q}$$\rightarrow$$0$ limit, where we find a strong enhancement of the magnetic component relative to the charge part.

In conclusion, we present a first-principles treatment of the generalized spin-charge density response tensor at the Tl/Si(111) surface. Our calculations demonstrate the appearance of a coupled spin-charge collective mode localized at the first few atomic layers close to the surface, which, as a direct consequence of the chiral spin texture of the Fermi contour, is composed by a transverse-spin density oscillation in addition to the usual charge density oscillation.
We resolve the real space details of this collective mode and show that the order of magnitude of both amplitudes is similar except for the small ${\bf q}$ limit, where the spin component is strongly enhanced with respect to the charge part.
Moreover, we show that this relative increase of the spin character should be understood as a general phenomenon, as long as the relevant electron band structure is composed by at least two circularly spin polarized bands crossing at the $\bar{\Gamma}$ point. 
The \textit{ab initio} character of our approach allows to explore other surface systems with more complex Fermi surfaces and spin textures, and paves the way to study -or even find- novel types of collective spin-charge excitations.


The authors acknowledge the Department of Education, Universities and Research of the Basque Government and UPV/EHU (Grant No. IT756-13) and the Spanish Ministry of Economy and Competitiveness MINECO (Grant No.  FIS2013-48286-C2-1-P and FIS2016-75862-P) for financial support. Computer facilities  were provided by the Donostia International Physics Center (DIPC). J.L. acknowledges DIPC and UPV/EHU (Grant No. PIF/UPV/16/240) for financial support.

\bibliography{bibliography}

\end{document}